\documentclass[12pt,a4paper]{iopart}
\usepackage{graphicx,iopams}  
\begin{document}
\bibliographystyle{unsrt}  %{iopart-num}

\title[Error Field Assesment from Driven Mode Rotation]
{Error Field Assessment from Driven Rotation of 
Stable External Kinks at EXTRAP-T2R Reversed Field Pinch}

\author{F.A.~Volpe$^1$, 
L.~Frassinetti$^2$, P.R.~Brunsell$^2$, J.R.~Drake$^2$, K.E.J.~Olofsson$^2$}
\address{$^1$ Columbia University, New York, NY, USA}
\address{$^2$ Royal Institute of Technology (KTH), Stockholm, Sweden}
\ead{fvolpe@columbia.edu}

\begin{abstract}
A new non-disruptive error field (EF) assessment technique not
restricted to low density and thus low beta was demonstrated at the
EXTRAP-T2R reversed field pinch. Stable and marginally stable 
external kink modes of toroidal mode number $n$=10 and $n$=8, 
respectively, were generated,
and their rotation sustained, by means of rotating magnetic
perturbations of the same $n$. Due
to finite EFs, and in spite of the applied perturbations rotating
uniformly and having constant amplitude, the kink modes were observed
to rotate non-uniformly and be modulated in amplitude. 
This behavior was used to precisely infer the amplitude and
approximately estimate the toroidal phase of the EF. A subsequent scan
permitted to optimize the toroidal phase. The technique was
tested against deliberately applied as well as intrinsic error
fields of $n$=8 and 10. Corrections equal and opposite to the estimated
error fields were applied. The efficacy of the error 
compensation was indicated by the increased discharge duration and 
more uniform mode rotation in response to a uniformly rotating
perturbation. The results are in good agreement with theory, and 
the extension to lower $n$, to 
tearing modes and to tokamaks, including ITER, is discussed. 

\end{abstract}

\submitto{Nuclear Fusion}
%\maketitle

%===========================================================================
\section{Introduction}

Error Fields (EFs) \cite{Reimerdes} are known to lower confinement in
toroidal  plasmas and, in general, to lower their rotation. 
An exception is the Neoclassical Toroidal Viscosity induced ``offset rotation'' 
\cite{ACole}. Additionally, EFs can seed magnetic
islands by EF penetration or cause locking of pre-existing islands
\cite{Fitzpatrick, Nave}, often resulting in disruptions \cite{DeVries}.
Correction techniques were developed to minimize EFs and their effects
\cite{Reimerdes}. The most widespread technique consists in reducing
the  density until a locked mode forms, with the achievement of lower
densities  indicating better EF correction: performing the ramp in
presence of different, deliberately applied static magnetic perturbations
(MPs), allows  to indirectly infer the EF and its best
correction. This, however, is the  best correction at low density and
thus at low beta,  but experimental evidence \cite{Schafer}  and
recent ITER modeling \cite{Park} suggest that the best correction  at
high beta can be significantly different due to plasma response
\cite{Park}. Furthermore, the technique described above needs several
discharges, each terminating with a locked mode and, often, a disruption. 

Here we present an EF correction (EFC) technique not based on the appearance 
of low-density Locked Modes, not restricted to low beta, 
non destructive, and requiring only a fraction of a discharge. In fact, 
it could be deployed multiple times within the same discharge, 
to dynamically assess and correct the EF in real time as it evolves. 

The technique was inspired by DIII-D experiments
in which rotating MPs  were used to rotate and reposition locked tearing 
modes of poloidal/toroidal mode number $m/n$=2/1 
and bring them in view of gyrotron launchers, for their
stabilization \cite{Volpe}. 
In spite of the MPs rotating uniformly, 
the experiments showed non-uniformities in the mode rotation. These 
observations were  
explained with the fact that rotation was taking place in the
presence of EFs, and the mode  was in fact locked to the resultant of
the static EF and the applied  rotating MP. Such resultant does not
rotate uniformly, neither is it constant in amplitude. In fact, if
the MP is weaker than the EF, the resultant doesn't even describe complete
revolutions and, for short periods of time, it can even rotate opposite 
to the MP \cite{Volpe}. 

The idea behind the present work is that (1) the applied MP is known, 
(2) the EF+MP resultant can be measured from the dynamics of a mode 
(not necessarily a tearing mode) 
locked to it and therefore (3) the EF can be deduced from the difference. 
Specifically, we applied uniformly rotating MPs and analyzed
the non-uniform mode rotation and the variations of mode amplitude to
precisely infer the amplitude and approximately estimate the toroidal
phase of the EF. A subsequent scan permitted to optimize the toroidal
phase.  The experiments were carried out at the  EXTRAP-T2R reversed
field pinch \cite{Brunsell_unstableRWMs} and were made possible by the coil control
system briefly described in Sec.~2. For simplicity, and unlike the
earlier DIII-D experiments with $m$=2, $n$=1 locked tearing modes, 
the modes used here were external kink modes of $m$=1 and 
high toroidal numbers, $n$=8 and 10. The rationale for this
decision, explained in Sec.~3, is that at EXTRAP-T2R these modes are
stable, which made the experiment easier.  Sec.~4 contains numerical
predictions of the time evolution of the  amplitude and toroidal phase
of a stable kink mode rotating in the presence  of a resistive wall,
static EFs and a rotating MP. The predicitions are based on a time-dependent 
cylindrical MHD model for thin-wall boundary conditions \cite{Gimblett} and  
are in good qualitative and quantitative agreement  with the 
experimental results presented in Sec.~5.  The transition from
incomplete to complete mode rotation allowed to  quantify the strength
of the $n$=8 and 10 EFs. Corrections of equal strength and varying
toroidal phase were applied by means of static MPs until the
toroidal phase was optimized. In this optimization we used two indicators 
of good EFC: the extended discharge duration and a more uniform mode rotation 
in response to a uniformly rotating perturbation.

%===========================================================================
\section{Experimental Set-up}   

EXTRAP-T2R \cite{Brunsell_unstableRWMs} is a reversed field pinch (RFP) 
of major radius $R$=1.24m  and plasma minor radius $a$=0.183m.  
External to the vessel is a double-layer Copper shell of thickness
$d$=2$\times$0.5mm.  A magnetic field of poloidal mode number $m$ and
toroidal mode number $n$ penetrates across this resistive shell (or
wall) in a time $\tau_{mn}$ which is a fraction of the ``long''
resistive wall time $\tau_w$. Theoretically, in the cylindrical
limit, this evaluates $\tau_w =\mu_0 \sigma r_w d$=13.6ms, where
$\sigma$ and $r_w$ are the conductivity and radius of the wall.
Experimentally, $\tau_w$ was measured to evaluate approximately
$\tau_w$=11.2ms, slightly larger than earlier estimates \cite{lf1}. 

Earlier works \cite{Brunsell,Drake_RWMs} provided evidence of 
EFs at various $n$. 
EFs at $n$=+8 and $n$=+10, which the present paper focuses on, 
might be due to the vacuum vessel 
being made of bellows sections joined with short flat sections 
(15 in total, placed at irregular toroidal angles). 

\begin{figure}[h]
  \centering
  \includegraphics{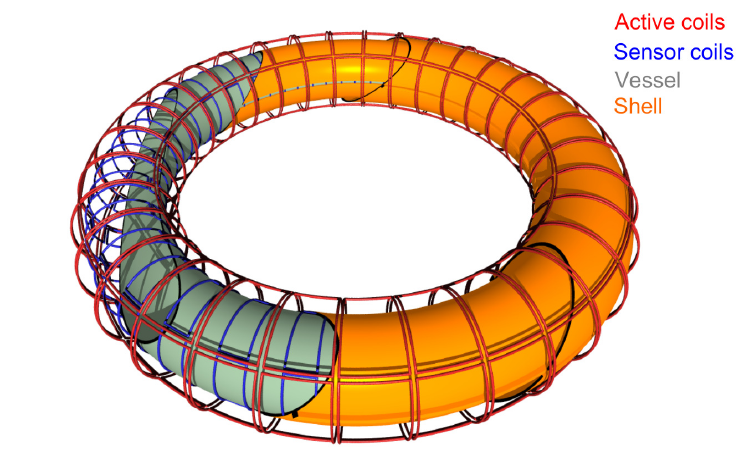}
  \caption{Schematic of (from inner to outer) vacuum vessel, 
  sensor coils, resistive shell and control coils at EXTRAP-T2R.}
  \label{Fig_EXTRAP}
\end{figure}

EXTRAP-T2R is equipped with a system for the active control of
resistive wall modes (RWMs) and EFs. External to the vessel, and just inside
the shell, is an array of  4 (poloidal) $\times$ 32 (toroidal)  saddle
loops used to sense the radial field,  depicted in
Fig.\ref{Fig_EXTRAP}. Located outside the shell is an array of
4$\times$32 control coils or actuators,  powered by suitably modified
audio amplifiers, used to apply radial field perturbations.  Various
control algorithms were developed and tested with success
\cite{ErikThesis,Frassinetti11}, but  of particular relevance here is 
a feedback algorithm presented in \cite{Erik}, 
capable of simultaneously sustaining preset finite
amplitudes  for all modes of $m$=1 and $-16 \leq n \leq 15$.
These amplitudes are typically set to 0, to suppress RWMs and EFs. 

The algorithm was modified for the sake of the experiments presented
here:  the capability was added to suppress all modes except 1-4
modes of up to 4 different $n$. Those unsuppressed modes were not
controlled in feedback. For the corresponding
values of $n$, it was possible to apply  static MPs of given amplitude
and toroidal phase and/or rotating MPs of given amplitude, frequency
and phase. In particular, rotating MPs were used to excite stable external 
kink modes and 
drive their rotation. On occasions, static MP were used to deliberately 
apply ``proxy'' EFs, stronger than the intrinsic machine EFs, 
thus easier to detect and optimal to test the method validity.

%===========================================================================
\section{Discussion on Modes and Mode Numbers}

The RFP equilibrium used for this experiment is defined by the reversal 
parameter \cite{Ortolani} $F=B_{\phi}(a)/ \langle B_{\phi} \rangle
=-0.25$   and the pinch parameter $\Theta=B_{\theta}(a)/ \langle B_{\phi} \rangle
$=1.66, where $B_\theta$ is the poloidal field and 
$\langle B_\phi \rangle$ is the toroidal field $B_\phi$ 
averaged over the plasma cross-section. 
Fig.~\ref{Fig_gamma}a summarizes the stability (growth rate $\gamma <$0) 
and instability ($\gamma >$0) of $m$=1 modes of various toroidal 
numbers $n$. The diagram was computed for the equilibrium used. 
However, small changes of equilibrium can have significant effects, thus the 
value of the diagram is mostly illustrative. 
Modes of $m \neq$1 are typically stable for the 
RFP equilibrium used (low $\Theta$) and will not be considered here. 
The convention on the sign of $n$ is as follows:  
modes with a handedness corresponding to
the pitch of field lines inside the reversal surface have $n <$ 0 and
modes outside the reversal surface have $n >$0. The reversal surface is defined 
as the surface where $B_\phi$ changes sign, located at approximately 
$r/a$=0.82.

Stable modes can only exist in presence of finite perturbations   
of the same $m$ and $n$. They form (disappear) on a timescale $\gamma ^{-1}$  
as soon as those perturbations are applied (removed). 

Note that ``modes'' here can refer to tearing or kink modes, depending 
whether the corresponding rational surface (resonance) is internal to the 
plasma and an island can form, or not. If external to the plasma, 
the perturbation can only kink the plasma itself. 
Very high positive $n$, though, can resonate at a surface external to the 
reversal surface, but inside the plasma. 

From left to right in Fig.\ref{Fig_gamma} we have: 

\begin{enumerate}
\item rapidly rotating (tens of kHz) Tearing Modes (TMs) of $n \lesssim $12. 
  These modes are internal, resonant ($q=-m/n$ in the plasma) and stable 
  \cite{Frassinetti_TMs}.   
\item unstable and marginally unstable kink modes ($-11 \lesssim n \lesssim 6$, 
  except $n=-1$, which calculations suggest being marginally stable) growing 
  on a time scale of the order of $\tau_w$ or slower. These are basically RWMs 
  \cite{Brunsell_unstableRWMs}.
\item stable kink modes ($n \gtrsim 7$) only exisiting in presence of an MP 
  of the same $n$. These mode grow (decay) on timescales of order 
  $\tau_w$ or slower as soon as the MP is turned on (off). However, they do not 
  grow indefinitely in time: they only grow up to a saturated value 
  proportional to the MP strength. The proportionality factor, or gain, is 
  dictated by the plasma response. These modes can also  be referred to as 
  stable RWMs \cite{Drake_RWMs}. 
\end{enumerate}

Correspondingly, the following considerations can be made with regard 
to suitability to EF assessment by driven rotation:

\begin{enumerate}
\item
  Shielding by the shell makes it impossible to couple rotating MPs 
  to fast-rotating modes 
  like the TMs mentioned above. Besides, previous work showed that, at 
  EXTRAP-T2R, the main 
  effect of static MPs and EFs on TMs is their magnetic braking 
  \cite{Frassinetti_TMs}. MPs rotating at few tens of Hz are expected 
  to have the same effect. 
\item 
  Unstable and marginally unstable RWMs can in principle be used, but their 
  growth would be in competition with other effects on their amplitude,  
  making it difficult to extract the effect of the EF. 
  These modes will be the subject of future work. 
\item
  RWMs of $n\ge 8$, however, are stable. This simplifies the extraction of 
  EF and MP effects on the RWM amplitude, compared with unstable RWMs of lower 
  $n$ mentioned above. 
  Also, being stable, they decay as soon as their MP+EF drive 
  is zeroed, which is also an indication of good EFC. 
  More generally, as soon as the drive is reduced or steered, so is the RWM. 
  Finally, stable RWMs 
  amplify EFs, making it easier to measure them with sensor coils. 
\end{enumerate}

It should also be mentioned that a suitable mode for 
the technique proposed  
% EF assessment by the technique outlined in the Introduction (Sec.1) 
does {\em not} need to be preexistent: 
it can be generated when necessary, for instance 
by a $\beta$ ramp at low 
rotation, like NTMs in the DIII-D experiments cited above \cite{Volpe}. 
It does not need to be 
rotating either. For example, it can be a non-rotating locked Tearing Mode
not preceded by a rotating precursor, but rather seeded by EF or MP 
penetration. On the other hand, it does not need to be exactly 
locked either: a slowly rotating Quasi-stationay Mode will still 
be affected by EFs in its rotation. 

Rather, the mode of choice needs to interact with the EF. The EF and, 
as a result, the mode, needs to depend on the toroidal angle $\phi$, and it 
must be possible to define a $\phi$-dependent potential energy, 
reaching a minimum for a certain value of $\phi$. For this, 
it is not necessary for the mode and the dominant EF to have the same 
$m$ and $n$: the interaction of modes and EFs of different mode numbers 
will still depend on $\phi$.

%===========================================================================
\section{Numerical Predictions}     \label{Sec_NumPredict}

Let us expand the radial field $b_r$ measured at the wall 
in poloidal and toroidal components 
$b_r^{m,n}$ in the cylindrical (large aspect ratio) limit: 

\begin{equation}
  b_r(r,\theta,\phi,t)=\sum_{m,n} b_r^{m,n}(r,t) \exp (im\theta +in\phi)
\end{equation}

where $r$, $\theta$ and $\phi$ are the radial, poloidal and toroidal coordinate 
respectively. 

In the absence of perturbations and control the growth rate $\gamma^{m,n}$ 
of a RWM of mode numbers $m$ and $n$ is defined by  

\begin{equation}
  \frac{\partial}{\partial t} b_r^{m,n}-\gamma^{m,n} b_r^{m,n}=0
  \label{Eq_gamma_def}.
\end{equation}

In presence of a thin shell (thin compared with the skin depth for the 
mode and timescale of interest), $\gamma^{m,n}$ 
is related to $\tau_w \simeq $11.2ms by \cite{Fitzpatrick}

\begin{equation}
  \label{Eq_gamma}
  \gamma^{m,n} \tau_w 
  =\frac{1}{b_r^{m,n}(r_w)}
  \left[ \frac{\partial (r b_r^{m,n})}{\partial r}\right]_{r_w^-}^{r_w^+}.
\end{equation}

The quantity in square brackets represents the jump of the derivative 
across the thin shell and 
$r_w^-$ and $r_w^+$ denote radial locations just inside and 
outside the radial location of the wall (and, with good approximation, of the 
sensors), $r_w$. 

In presence of a time-dependent external perturbation $b_{ext}^{m,n}$, 
Eq.~\ref{Eq_gamma_def}, describing the time-evolution of the radial-field 
harmonics evaluated at the wall, $b_r^{m,n}$, modifies as follows 
\cite{Fitzpatrick,Gimblett,Newcomb,Pustovitov_B,Pustovitov_FA,Pustovitov_Fiz,Drake,Brunsell,Gregoratto}: 

\begin{equation}
  \frac{\partial}{\partial t} b_r^{m,n} - 
  \gamma^{m,n} b_r^{m,n}=\frac{b_{ext}^{m,n}}{\tau^{m,n}},
  \label{Eq_bwall}
\end{equation}

where we adopted the notations of \cite{Drake,Brunsell,Gregoratto}, and   
$\tau^{m,n}$ is the wall diffusion time for $b_{ext}^{m,n}$, related to $\tau_w$ by 

\begin{equation}
 \label{Eq_taumn}
 \frac{\tau_w}{\tau^{m,n}}=
 \frac{1}{b_r^v(r_w)}
 \left[ \frac{\partial (r b_r^v)}{\partial r}\right]_{r_w^-}^{r_w^+},
\end{equation}

Here $b_r^v$ is the radial component of the vacuum field, i.e.~in 
absence of plasma, but in presence of the thin resistive wall. 
The jump in the radial derivative across the shell  
on the right hand side of the equation is due to 
eddy currents induced in the conductive thin shell itself. 
These eddy currents can be due to modes in the plasma and/or currents in the 
control coils. 

\begin{figure}[h]
  \centering
  \includegraphics{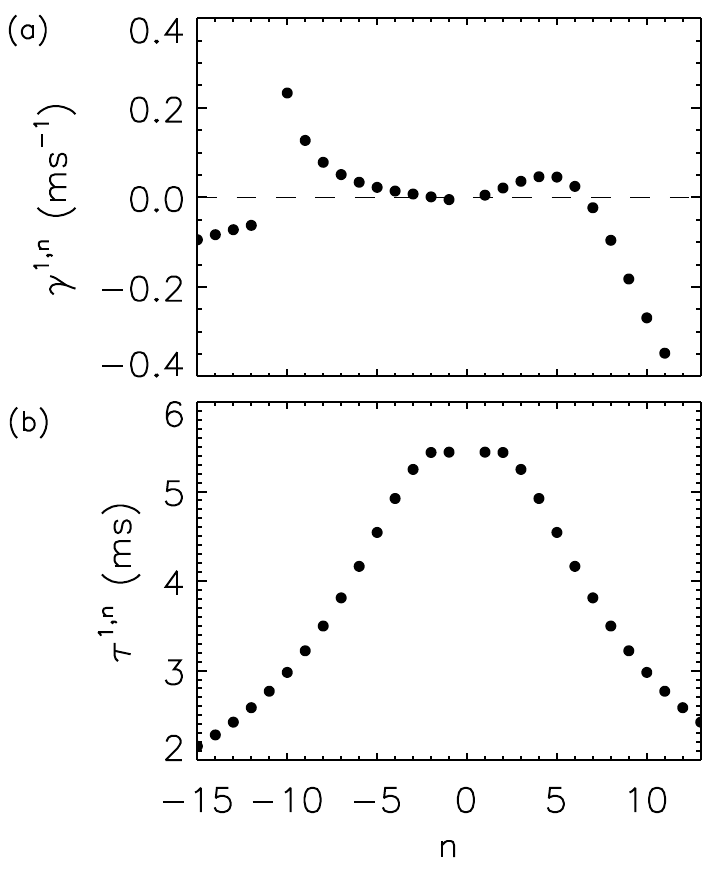}
  \caption{(a) $m$=1 RWM growth rate and (b) wall diffusion time 
      as a function of $n$ for the equilibrium used in the present work.}
  \label{Fig_gamma}
\end{figure}

The $n$ dependence of $\gamma^{m,n}$ and $\tau^{m,n}$ in EXTRAP T2R for $m$=1 
for the plasma equilibrium used in the present work are plotted in 
Fig.\ref{Fig_gamma}.

In the remainder, for simplicity, superscripts $^{m,n}$ will be omitted from  
$b_r^{m,n}$, $b_{ext}^{m,n}$, $\tau^{m,n}$ and $\gamma^{m,n}$. 
For our purposes $b_{ext}$ consists of a static field of 
amplitude $b_0$ (the resultant of the intrinsic EF, proxy EF, if any, and  
deliberately applied static MP, if any)   
and a rotating field of amplitude $b_1$:

\begin{equation}
  b_{ext}(t)=b_0+b_1 e^{i\omega t},
  \label{Eq_bext}
\end{equation}

where $\omega$ is the angular frequency at which the field, or equivalently  
the currents in the active coils, oscillate. 
A mode of toroidal number $n$ rotates with angular velocity $\omega/n$. 

Note that both $b_0$ and $b_1$ are complex and determine the phase of 
$b_{ext}(t)$ and that Eq.~\ref{Eq_bwall} can be analytically solved. 
Assuming as initial condition $b_r(0)=0$, we obtain: 

\begin{equation}
 b_r= \frac{b_0}{\gamma \tau} 
  \left(e^{\gamma t}-1\right)+\frac{b_1}{(\gamma-i\omega)\tau}
  \left( e^{\gamma t}-e^{i\omega t} \right)
 \label{Eq_bsim}
\end{equation}

This expression describes the behavior of the radial magnetic field
perturbation measured at the sensors. This measured field 
(Eq.\ref{Eq_bsim}) is due to the applied external perturbation 
(Eq.\ref{Eq_bext}) but is modified by the plasma response. 

\begin{figure}[!ht]
    \centerline{\includegraphics{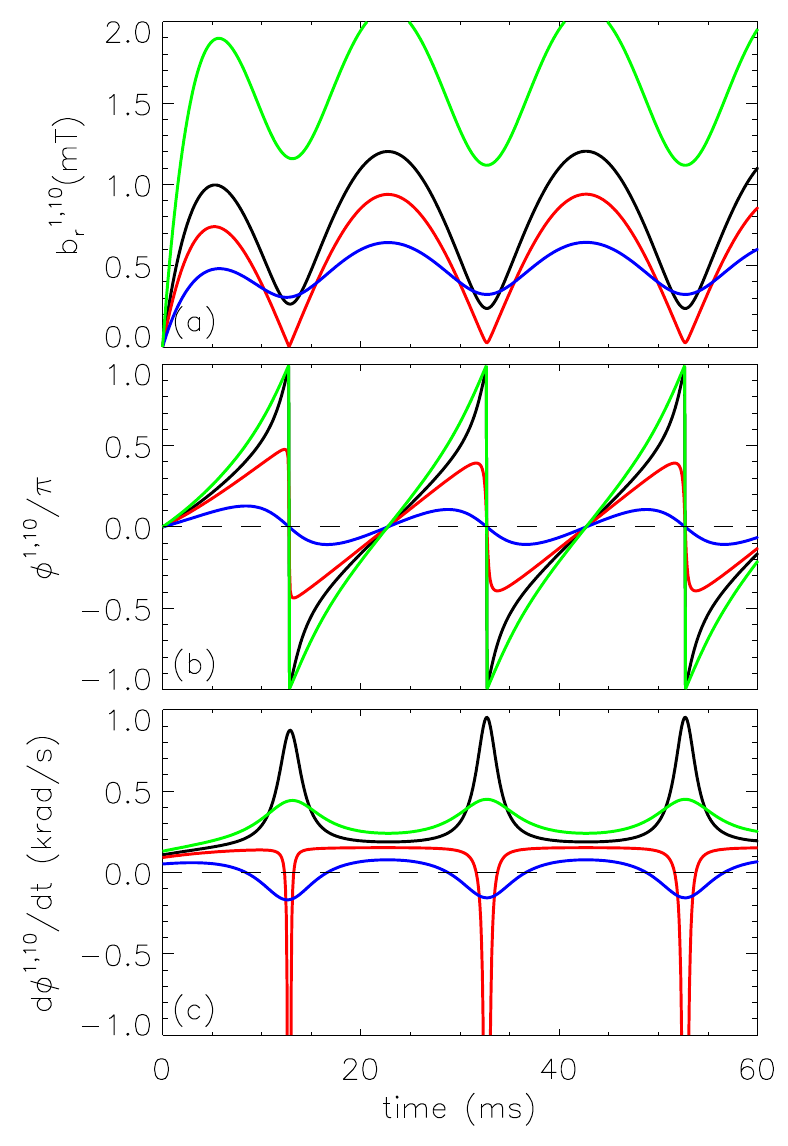}}
    \caption{Calculated (a) amplitude, (b) phase and (c) angular velocity of  
      $m$=1, $n$=10 component of radial field at the sensor coils, 
      modeled after Eq.\ref{Eq_bsim}, for EF amplitude $b_0=0.4mT$ and 
      MP rotating frequency $\omega/2\pi$=50Hz. 
      The amplitudes of the rotating perturbations are $b_1=0.2$mT (blue
      curves), $b_1=0.6$mT (red), $b_1=0.9$mT (black) and $b_1=2$mT (green). 
      Note that the static EF evaluates 0.4mT, but the transition occurs at 
      a rotating field amplitude 0.6mT$<b_1<$0.9mT.}
    \label{Fig_mod}
\end{figure}

An initial transient of order $\gamma^{-1}$ is 
visible in Fig.\ref{Fig_mod}. This transient typically lasts few ms 
(see plots of $\gamma$ in Fig.\ref{Fig_gamma}a,  
for various modes). After the transient, Eq.\ref{Eq_bsim} for stable modes 
($\gamma<$0) simplifies as follows:

\begin{equation}
  b_r = |b_r| e^{i\omega t+\Delta \phi}
 \label{Eq_bsimplif}
\end{equation}

where

\begin{eqnarray}
  |b_r| & = & \frac{b_1}{\tau \sqrt{\omega^2 +\gamma^2}} \label{Eq_b1}\\
  \Delta \phi &=& \arctan \frac{\omega}{\gamma}
\end{eqnarray}

Eq.\ref{Eq_bsim} is used to simulate the experimental
results with $\tau_w=11.2$ms and $\gamma^{m,n}$ and $\tau^{m,n}$ from
Fig.\ref{Fig_gamma}. 
Fig.\ref{Fig_mod} shows simulation results for $n$=10, 
a static field $b_0$=0.4mT and four values 
of the probing field $b_1$, rotating at 50Hz. 
Note that in the experiment $b_1$ is known, while $b_0$ is the 
unknown total static EF, which can be determined as follows. 

A small rotating field $b_1 \ll b_0 \tau \sqrt{\omega^2 +\gamma^2}$ 
does not drive a complete 
rotation of the mode, but only an oscillation around the phase of $b_0$ 
(here assumed 0), accompanied by large amplitude variations  
(blue curves in Fig.\ref{Fig_mod}). 
Under the effect of a large rotating field 
$b_1 \gg b_0 \tau \sqrt{\omega^2 +\gamma^2}$, instead, the 
sensors register a complete, uniform rotation and smaller relative changes of 
amplitude (green curves). 
Finally, when the rotationally shielded probing-field magnitude 
$\frac{b_1}{\tau \sqrt{\omega^2 +\gamma^2}}$ (Eq.\ref{Eq_b1})
is slightly smaller or larger 
than the EF $b_0$, the result is a large oscillation in the first case (red), 
and a complete, although highly non-uniform rotation in the second (black). 
Correspondingly the angular velocity has negative or positive peaks, 
respectively (Fig.\ref{Fig_mod}c). 
The transition from one behavior (oscillations) to the other (complete
rotations) occurs for $\frac{b_1}{\tau \sqrt{\omega^2 +\gamma^2}}=b_0$. 
Experimentally identifying this
transition  allows to indirectly measure the EF amplitude $b_0$.
Also, $b_0$ has toroidal phase $\phi$=0 in the example of
Fig.\ref{Fig_mod},  but this can be easily generalized: if its phase
is $\phi$, the  measured field will oscillate around $\phi$. 
$\phi$ can also be measured when $b_r$ reaches its minimum, i.e.~as 
the mid-point of the  large phase oscillation
recorded before the mode describes a complete  rotation. However, that is  a
large oscillation occurring in a short time, hence this measurement of
$\phi$ suffers from errors. 
In fact, the closer one gets to the transition, the preciser is the 
measurement of $|b_0|$, but the higher is $|d\phi/dt|$, 
thus the more rapid the variation of $\phi$ and 
the less precise its measurement, as if the two quantities 
$|b_0|$ and $\phi$ were related by an indetermination principle. 

\begin{figure}[t]
    \centerline{\includegraphics{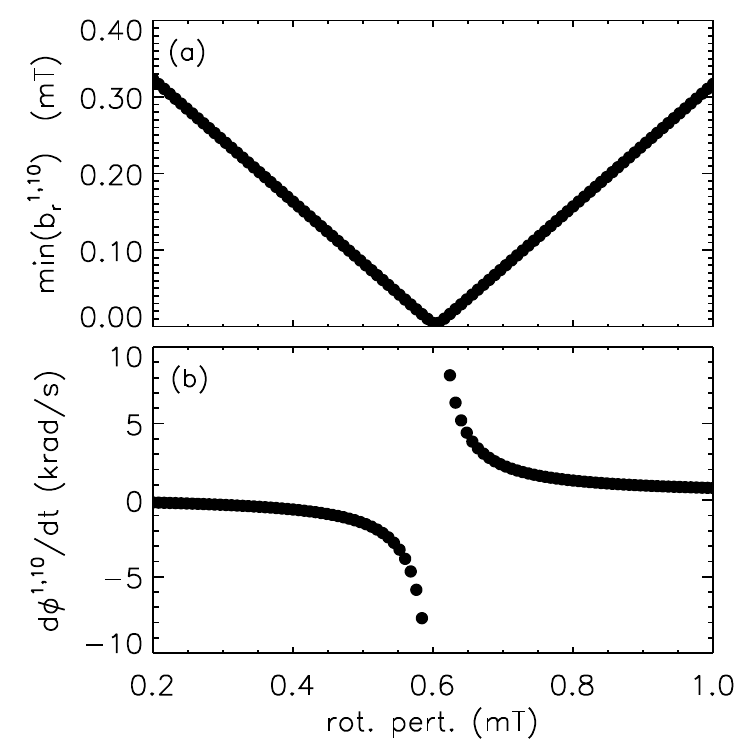}}
    \caption{(a) minimum amplitude of $b_r^{1,10}$ as calculated in 
      Fig.\ref{Fig_mod}a, as a function of the MP strength $b_1$ 
      for EF amplitude $b_0$=0.4mT and MP rotation frequency 50Hz. 
      (b) corresponding phase velocity at the time of minimum amplitude. 
      Error bars are smaller than symbol sizes.}
    \label{Fig_sum}
\end{figure}

Fig.\ref{Fig_mod} displays curves for four values of the rotating field 
$b_1$, but the calculations were repeated for several other values. The 
results are summarized in Fig.\ref{Fig_sum}, as a function of $b_1$. 
In particular, Fig.\ref{Fig_sum}a shows the minimum the minimum 
sensor measurement $b_r^{1,10}$ that 
can be reached at some time (i.e., for some toroidal orientation of the 
applied MP) for a given amplitude $b_1$ of the rotating probing field. 
Fig.\ref{Fig_sum}b presents 
the modeled phase velocity of the measured perturbation, $d\phi^{1,10}/dt$, 
reached at that same time. As already seen in 
Fig.\ref{Fig_mod}, when $b_r^{1,10}$ reaches a minimum, 
the angular velocity reaches 
an extreme (either a negative minimum, if the rotation is incomplete, or a 
positive maximum, if the rotation is complete). 

Perfect error field cancellation is obtained when
the modeled measured $b_r^{1,10}$ is suppressed to zero ($b_1\approx$0.6mT in 
Fig.\ref{Fig_sum}a). 
At that time, the angular speed $d\phi^{1,10}/dt$ 
transitions from strongly negative to strongly positive (Fig.\ref{Fig_sum}(b)). 

Note that observing a transition for a certain value of $b_1$ in 
Figs.\ref{Fig_mod}-\ref{Fig_sum} does not imply that the EF $b_0$ takes 
that same value. This would only be true in the limit of very slow rotation, 
$\omega\rightarrow 0$. Otherwise, the amplitude 
$b_1$ of the rotating perturbation (as applied outside 
the shell) undergoes some screening, and a stronger $b_1$ needs to be 
applied to cancel a given $b_0$.  
From Eq.\ref{Eq_b1}, we can estimate the screening factor to evaluate 
$1/\tau \sqrt{\omega^2 +\gamma^2} \approx 0.7$ for 
$\omega/2\pi$=50Hz. 
Thus, observing a transition at 
$b_1\simeq$0.6mT in Figs.\ref{Fig_mod}-\ref{Fig_sum} agrees with the fact that 
the EF assumed above for this example evaluates $b_0$=0.4mT.  

%The example in Figs.\ref{Fig_mod}-\ref{Fig_sum} 
%is indicative, is not meant to model the experimental results 
%presented below. $\omega$ is the same as in the experiments, 
%but $b_0$ and $b_1$ used for the example are not. 

%===========================================================================
\section{Experimental Results}

Here we present the proof of principle of the proposed EF assessment technique 
based on driven rotation of modes. 
For the reasons discussed in Sec.3, it was decided to use stable 
external kinks and characterize EFs of the same mode numbers: $m$=1, $n$=8 
and $m=1$, $n$=10. 
We started with $n$=10 kinks (more stable) and 
progressed to $n$=8, which was found to be the lowest stable $n$ and to have 
a smaller, thus, more difficult detect, intrinsic EF associated with it.

%All sensor measurements presented below were ``vacuum-subtracted'' to 
%remove the effect of the vertical and toroidal field, but not of the static or 
%rotating MP, on the sensor coils, in the following sense.   
%The vertical and toroidal field coils were energized 
%in dedicated ``vacuum shots'' in which no plasma was created. 
%Sensor-coil measurements in such vacuum shots 
%were then subtracted from the corresponding measurements in 
%actual plasma discharges. 
%Note that only the vertical and toroidal fields were subtracted. 

\subsection{Results for $n$=10}

\begin{figure}[t]
    \centerline{\includegraphics{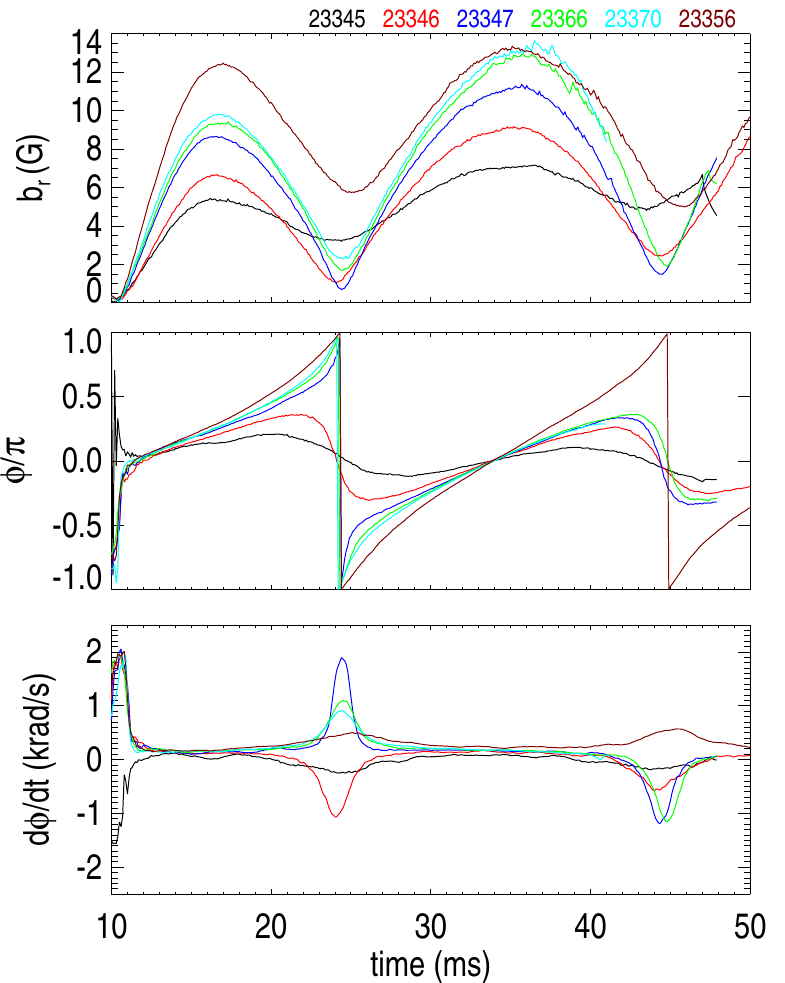}}
    \caption{Experimental (a) amplitude, (b) phase and (c) 1ms-smoothed 
      angular velocity of 
      $m$=1, $n$=10 component of radial field at the sensor coils, in 
      presence of ``proxy'' EF of amplitude $b_0=0.4mT$ and phase $\phi$=0 
      (same as n Fig.\ref{Fig_mod}).  
      MPs rotate at $\omega/2\pi$=50Hz, and have various amplitudes 
      (see Fig.\ref{Fig_n10LProxyAn}). Transition to complete rotation  
      occurs between $b_1=$0.35 mT and $b_1=$0.55 mT. 
      Signals are only plotted for the 
      duration of the discharges. Error bars are comparable with or 
      smaller than symbol sizes.}
     \label{Fig_n10LProxyRaw}
\end{figure}

For the first test,  
static MPs were used to apply ``proxy'' EFs of $n$=10, i.e.~fictitious EFs, 
stronger than usual and thus easier to detect. 
The $n$=10 EF was suppressed by  
automatic feedback as usual \cite{Erik,ErikThesis,Frassinetti11}  
in the first 10ms of the discharges illustrated in Fig.\ref{Fig_n10LProxyRaw}. 
At $t \ge$10ms, though, the intrinsic $n$=10 EF was left uncontrolled and an 
MP was applied, to simulate an $n$=10 EF of 0.92mT. 
In order for such EF to be constant, the actual programmed MP applied 
was increasing slightly on a timescale comparable with the discharge duration. 
This timescale is unrelated to $\omega^{-1}$, $\gamma^{-1}$, 
$\tau_w$ or $\tau^{m,n}$,  
but rather due to subtleties in the control hardware \cite{lf1}. 

\begin{figure}[ht]
    \centerline{\includegraphics{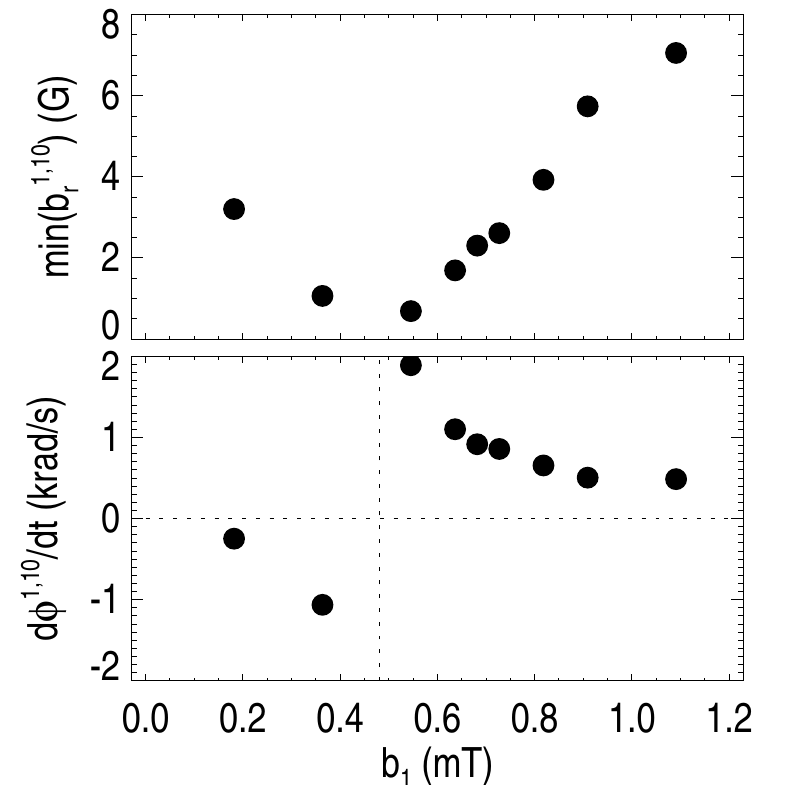}}
    \caption{(a) minimum amplitude of $b_r^{1,10}$ evaluated from 
      Fig.\ref{Fig_n10LProxyRaw}a and alike, as a function of the MP strength 
      $b_1$ for EF amplitude $b_0$=0.92 mT and MP rotation frequency 50Hz 
      (same values as in Fig.\ref{Fig_sum}). 
      (b) corresponding phase velocity at the time of minimum amplitude.}
    \label{Fig_n10LProxyAn}
\end{figure}

In the same time-interval ($t \ge$10ms) an $n$=10 rotating MP was also applied. 
Its rotation frequency, 50Hz, enables the observation of at least 
two rotation periods within a typical EXTRAP-T2R discharge. 
The resultant of the static MP, rotating MPs, and of the much smaller   
uncorrected EF, excites a RWM of $n$=10 and thus of 
negative growth rate (Fig.\ref{Fig_gamma}a), that is, stable. 
The time-evolution of its amplitude, toroidal phase and angular 
velocity is regulated by Eq.\ref{Eq_bsim} or, after a transient of few ms, 
by its approximation, Eq.\ref{Eq_bsimplif}. 
Indeed we obtained sensor coil measurements 
(Figs.\ref{Fig_n10LProxyRaw}-\ref{Fig_n10LProxyAn}) 
very similar to the calculations in Figs.\ref{Fig_mod}-\ref{Fig_sum}, 
including the hyperbolic behavior of $d\phi^{1,10}/dt$. 
In particular, the transition from incomplete to complete mode rotation 
occurs between $b_1$=0.35mT and $b_1$=0.55mT (Fig.\ref{Fig_n10LProxyRaw}b), 
at $b_1=0.48\pm0.05$mT (Fig.\ref{Fig_n10LProxyAn}b). 
For consistency and comparison with Eq.\ref{Eq_bsim}, we 
retained in the sensor measurements 
the direct contribution of the applied MPs (both static and rotating). 
In other types of analysis, e.g.~to extract the field $b_r$ associated with a 
mode, the contribution of the MPs could have been removed by ``vacuum 
subtraction'', i.e.~by taking separate measurements in presence of MPs but in 
absence of plasma, and subtracting them from the actual plasma measurements. 
This was not the case here, because $b_0$ and $b_1$ defined above  
include the applied static and rotating MPs. 
On the other hand, the effect of the vertical (VF) and toroidal field (TF) 
coils on the 
sensor measurements was vacuum-subtracted. The reason is that those coils 
introduce a high background in the $b_r$ measurements, but such background is 
nearly perfectly axisymmetric, i.e.~the VF and TF are not significant sources 
of $n$=10 EF. 

With the screening factor $1/\tau \sqrt{\omega^2 +\gamma^2}$  
taken into account, it is ``estimated'' that the proxy EF (which is 
actually known, but is treated as an unknown for this test) amounts to 
$b_0=0.89 \pm 0.10$mT. 
The error results from the propagation of the uncertainties in the 
measured screening factor (circa $\pm 10\%$) and in the rotating MP amplitude 
(while the programmed amplitude was constant, the actual amplitude was 
observed to fluctuate by approximately $\pm5\%$). 
The $b_0=0.89 \pm 0.10$mT measurement agrees with the applied field evaluating  
$b_0$=0.92mT although, rigorously, it needs to be compared with the EF+MP 
resultant, which is done below. 

The toroidal phase can be estimated from 
Fig.\ref{Fig_n10LProxyRaw}a to be $\phi \simeq (0\pm 0.1)\pi$, 
as this is the value around which signals oscillate. 
Again, this is in agreeement with the applied proxy having $\phi$=0. 
Alternatively, the EF phase could be inferred from the measurement of $\phi$ 
at the time when $b_r$ reaches its minimum i.e.~when 
the EF (the proxy EF) is best corrected by the EFC (the rotating MP). 
However, care should be exerted because $\phi$ varies rapidly in the vicinity 
of that optimal value (except of course if the rotating MP is strong enough to 
dominate the kink dynamics and make it rotate uniformly). 

The screening factor is $1/\tau \sqrt{\omega^2 +\gamma^2}$, and  
in Sec.\ref{Sec_NumPredict} it was calculated to evaluate approximately 0.7, 
on the basis of the prescribed $\omega$, 
calculated $\gamma$ (Fig.\ref{Fig_gamma}a) 
and calculated $\tau$ (Fig.\ref{Fig_gamma}b). 
In the experiment, however, rather than calculating it, we 
estimated the screening factor experimentally by comparing sensor measurements 
in two plasma discharges with, respectively, large static and large rotating  
MPs. The MPs were large in order for the intrinsic EF to be negligible. 
Comparing the two discharges showed that, for $m$=1 and $n$=10, rotating MPs of 
$\omega/2\pi$=50Hz generate, in presence of plasma, weaker signals 
(by a factor 0.54$\pm$0.05) than static MPs of the same intensity. 

\begin{figure}[ht]
    \centerline{\includegraphics{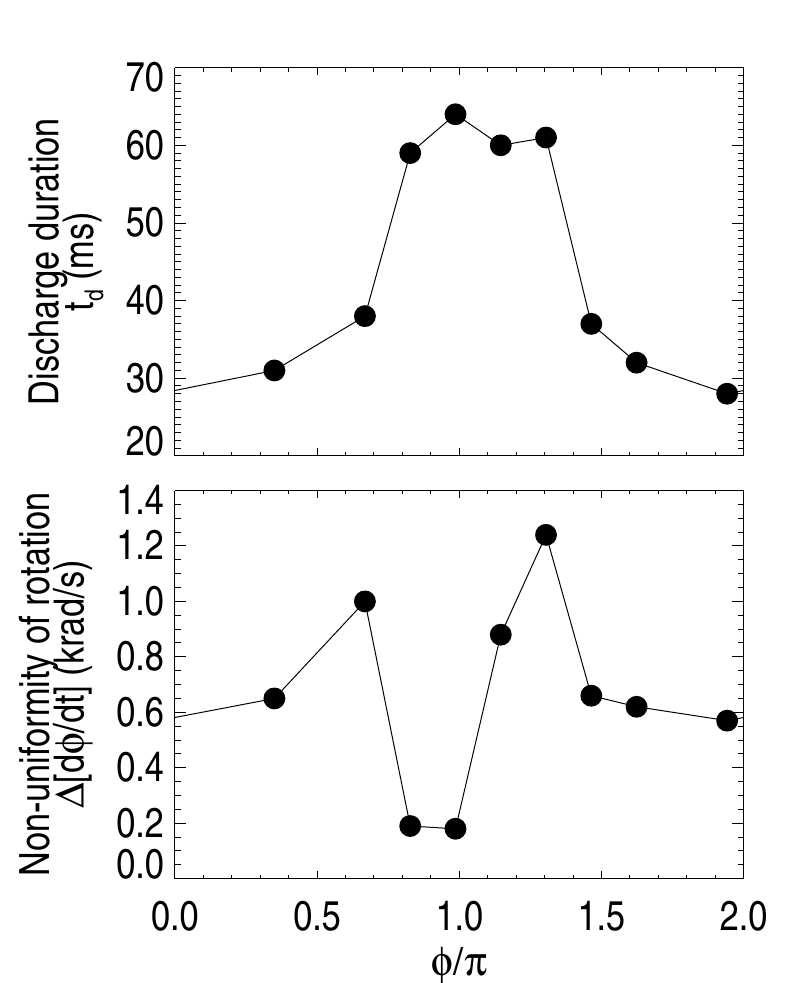}}
    \caption{Optimization of the toroidal phase $\phi$ of the $n$=10 EFC 
      and evidence of good EFC for optimal $\phi$, as evinced from (a) longer 
      duration of the discharge and (b) more uniform rotation.}
    \label{Fig_n10_PhiScan}
\end{figure}

For a precise determination of $\phi$, as well as to confirm the 
good quality of the EFC, corrections were applied, of amplitude 0.89mT and 
various $\phi$, in two sets of discharges, and two indicators of good EFC were 
monitored. 
The first indicator is the duration of the discharge: it is well-known that 
an improved EFC leads to better confinement and thus longer discharges, 
as indeed shown in Fig.\ref{Fig_n10_PhiScan}a. The broad maximum at 
$\phi = (1.05 \pm 0.25)\pi$ agrees with the expected best EFC having 
$\phi=\pi$. 

In another set of discharges, the EFC phase $\phi$ was scanned from one 
discharge to the other, a small, uniformly rotating MP was applied, and the 
uniformity of the measured mode rotation was studied: for perfect EFC, the 
mode was expected to rotate uniformly as well. Fig.\ref{Fig_n10_PhiScan}b 
confirms that the smaller variation of angular velocity during a 
rotation period is obtained for $\phi= (0.9 \pm 0.1)\pi$, 
in agreement with expectations. 

\begin{figure}[ht]
    \centerline{\includegraphics{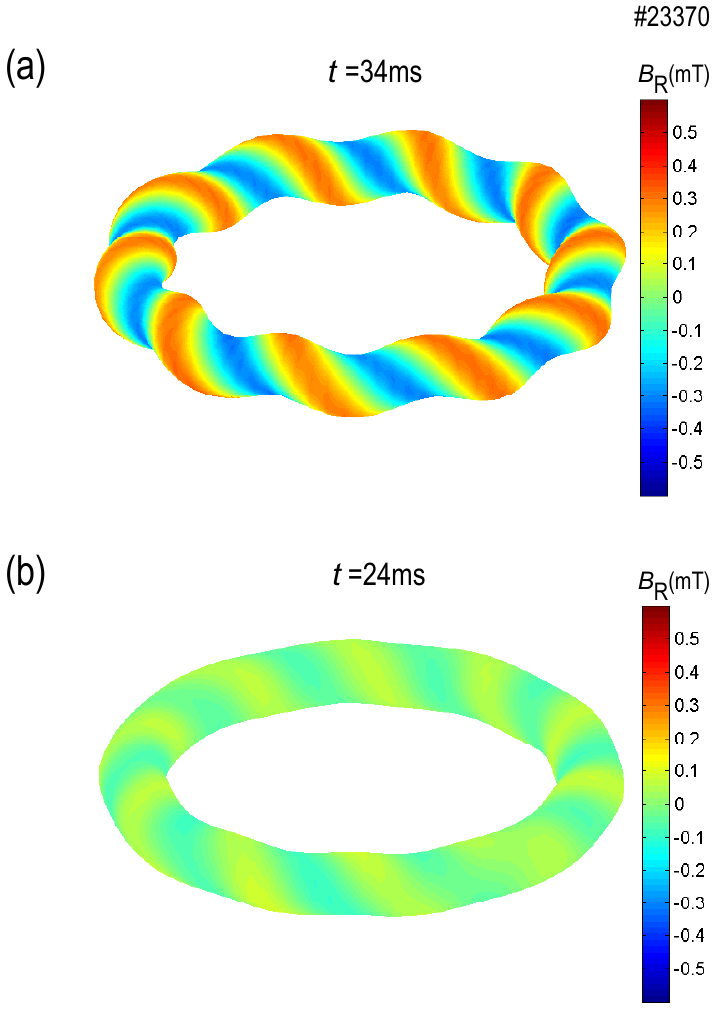}}
    \caption{Reconstruction of $m$=1, $n$=10 external kink mode (stable RWM) 
    based on actual saddle coil measurements at times when the MP (a) 
    reinforces or (b) nearly cancels the EF. For clarity the radial 
    deformation is exaggerated by a factor 10.}
    \label{Fig_cartoon_n10_mode}
\end{figure}

To further illustrate the good EFC, 
Fig.\ref{Fig_cartoon_n10_mode} shows a reconstruction of the plasma 
edge and radial field based on actual sensor data. Two features can be clearly 
recognized: the strong $m$=1, $n$=10 
perturbation at a time when the rotating MP has the same $\phi$ as 
the proxy EF, and its minimization half a rotation-period 
later (when the MP and proxy EF nearly cancel each other). 

\begin{figure}[t]
    \centerline{\includegraphics{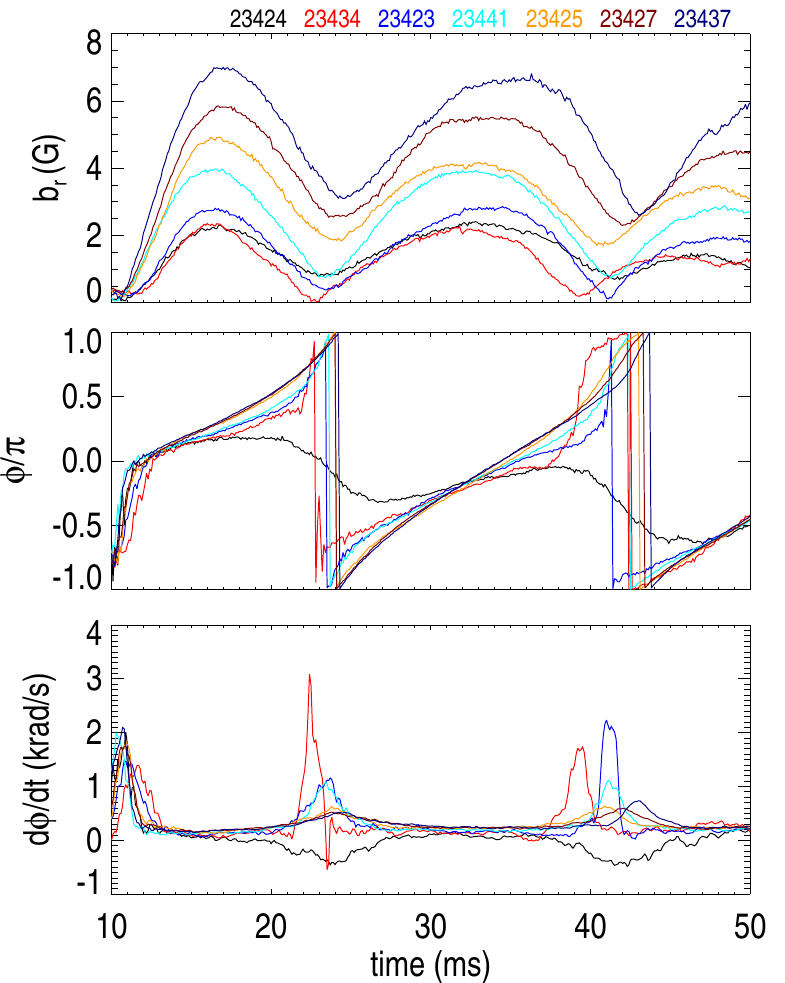}}
    \caption{Like Fig.\ref{Fig_n10LProxyRaw}, 
      except that no ``proxy'' EF is applied. Here the only 
      static $b_0$ is the intrinsic $m$=1, $n$=10 EF.  
      Transition to complete rotation occurs between $b_1=$0.078 mT 
      and $b_1=$0.116 mT.}
     \label{Fig_n10IntrinsRaw}
\end{figure}

\begin{figure}[ht]
    \centerline{\includegraphics{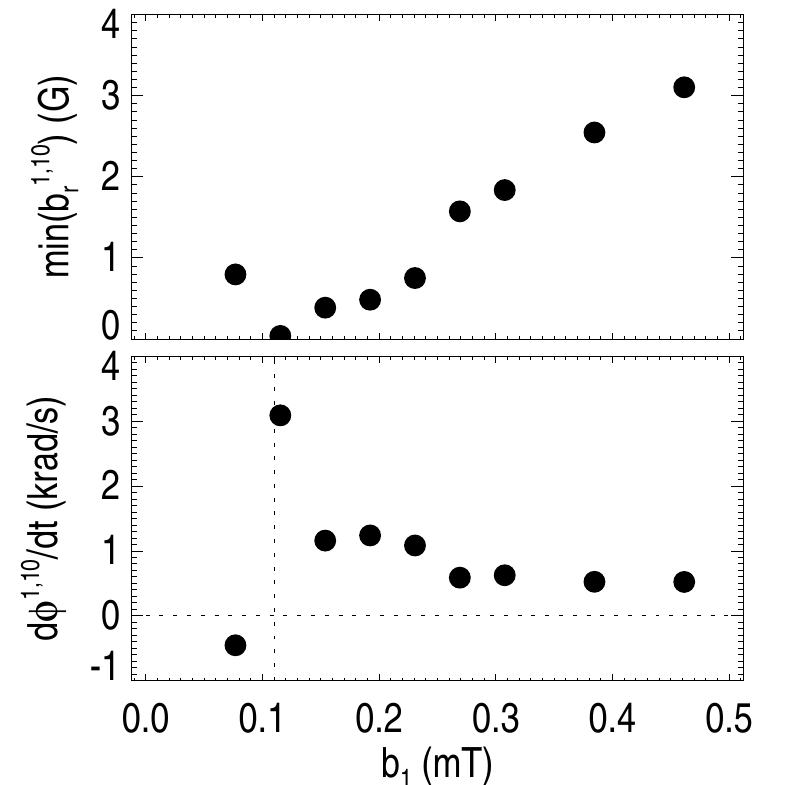}}
     \caption{Like Fig.\ref{Fig_n10LProxyAn},  
     but for the $n$=10 intrinsic EF results of Fig.\ref{Fig_n10IntrinsRaw}.}
    \label{Fig_n10IntrinsAn}
\end{figure}

After additional tests with a smaller proxy EF (0.23mT, not shown for 
brevity), proxy EFs were eventually removed and the 
the technique was applied to the assessment of the intrinsic $m$=1, $n$=10 EF. 
The results are presented in Fig.\ref{Fig_n10IntrinsRaw} and analyzed in 
Fig.\ref{Fig_n10IntrinsAn}.
The best correction is obtained for a rotating perturbation $b_1$=0.11mT, 
corresponding to a static (unshielded) $b_0=0.20\pm0.02$mT, 
in agreement with earlier estimates \cite{Brunsell,Drake_RWMs}. 
Fig.\ref{Fig_n10IntrinsRaw}b indicates that its toroidal phase is 
approximately $\phi$=0. 

In retrospect the total static field error EF+MP in the proxy experiment 
presented above amounted to $1.12 \pm 0.04$mT and had phase 
$\phi \simeq (0\pm 0.1)\pi$,  
agreeing within two standard deviations with the measurements, 
$b_0=0.89\pm0.10$mT and $\phi \simeq (0\pm 0.1)\pi$.

\subsection{Results for $n$=8}

\begin{figure}[t]
    \centerline{\includegraphics{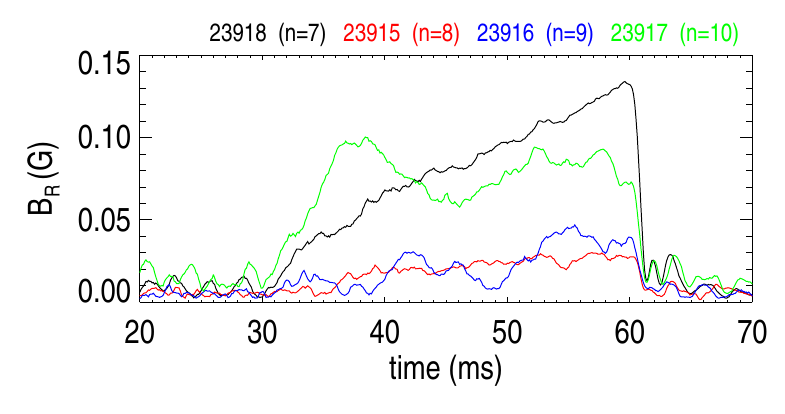}}
    \caption{Measured amplitude of $m$=1 modes of various $n$, in discharges 
    where the corresponding $n$ was left uncontrolled at $t$=30-60ms. 
    The $n$=7 signal, unstable, is due to a growing RWM destabilized by the 
    uncontrolled $n$=7 EF. Other signals ($n$=8, 9 and 10) take finite but 
    small, non-growing (stable) values, due to intrinsic EFs, possibly 
    amplified or attenuated by the plasma response. Note that both the EFs and 
    plasma response can fluctuate with time.}
     \label{Fig_n78910}
     %%%%%% n=7 23912,13,18; n=8 23910,14,15; n=9 23916; n=10 23917
\end{figure}

The $n$=10 results presented above are easily extended to $n>$10 at 
EXTRAP-T2R, as the corresponding RWMs are even more stable. 
For this reason, it is more interesting to extend the technique to lower $n$. 
The extension to unstable $n$ will be the subject of a future work. 
In the present Section, we concentrate on marginally stable modes. 
It was experimentally determined that the lowest positive stable $n$ is 
$n$=8 (Fig.\ref{Fig_n78910}), in rough agreement with the prediction in  
Fig.\ref{Fig_gamma}a, that modes of $n\gtrsim 7$ are stable. 

\begin{figure}[t]
    \centerline{\includegraphics{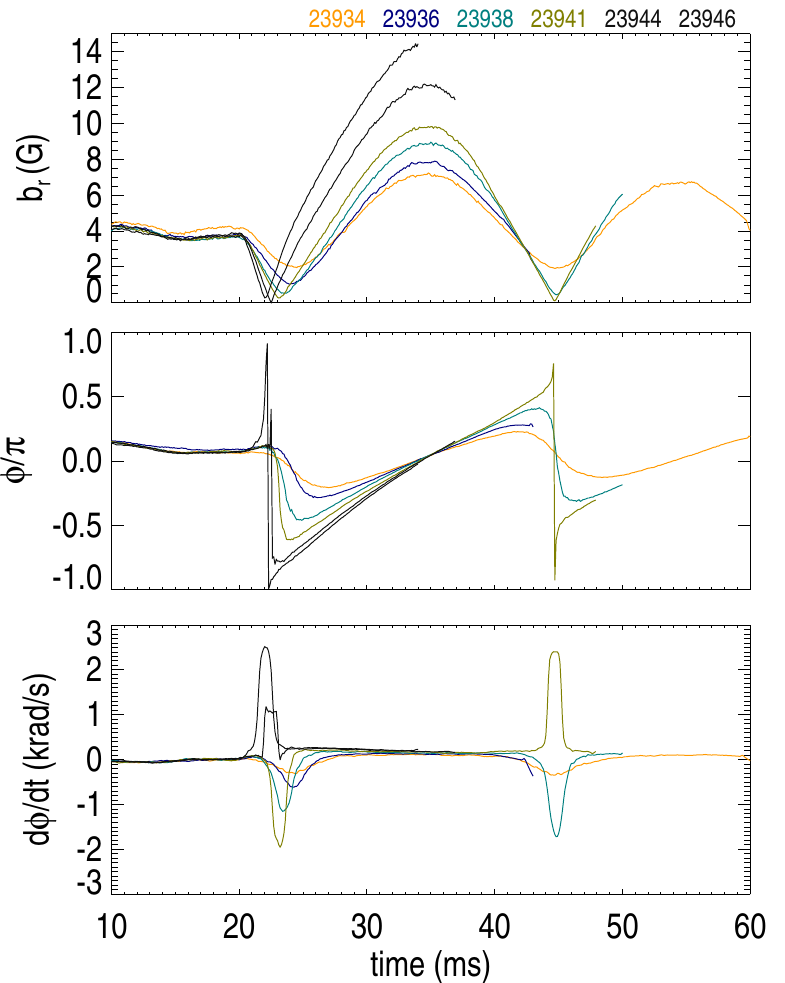}}
    \caption{Similar to Fig.\ref{Fig_n10LProxyRaw}, but for $n$=8. 
      Also, feedback is disabled and the static and rotating perturbations 
      applied at $t\ge$20ms instead of $t\ge$10ms. 
      Transition to complete rotation occurs between $b_1=$0.42mT and 
      $b_1=$0.46mT.}
     \label{Fig_n8LProxyRaw}
     %%%%%% SHOTS 23922-27 and 23934-47, except 940 and 945 (different proxy)
\end{figure}

\begin{figure}[ht]
    \centerline{\includegraphics{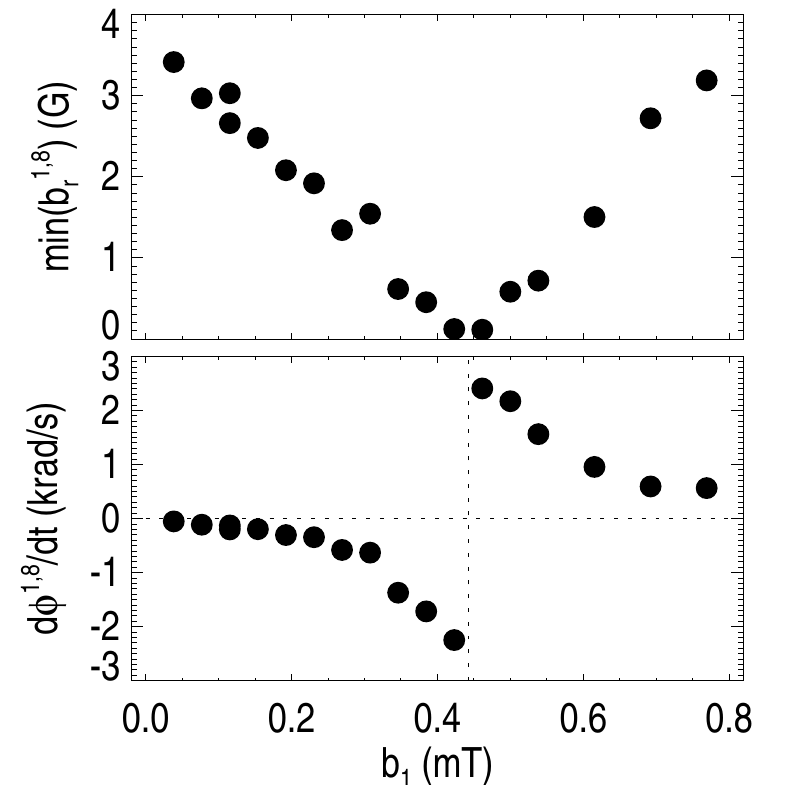}}
    \caption{Like Fig.\ref{Fig_n10LProxyAn}, but for the $n$=8 results of 
      Fig.\ref{Fig_n8LProxyRaw}.}
    \label{Fig_n8LProxyAn}
\end{figure}

Similar to the $n$=10 experiments,   
a probing $n$=8 MP rotating at 50Hz was applied, and its amplitude was  
varied from one discharge to the other. 
In a first set of discharges 
(Figs.\ref{Fig_n8LProxyRaw}-\ref{Fig_n8LProxyAn}), 
an $n$=8, $\phi$=0 proxy EF of 0.41 mT was applied.
The screening factor was $0.57\pm0.06$ and the transition to complete rotations 
occurred at $b_1$=0.44mT. This implies a $b_0=0.77\pm0.08$mT estimate 
which should not be compared with the static MP, but with 
the EF+MP resultant (see below). The toroidal phase is 
$\phi/\pi = 0.1 \pm 0.1$. 

\begin{figure}[t]
    \centerline{\includegraphics{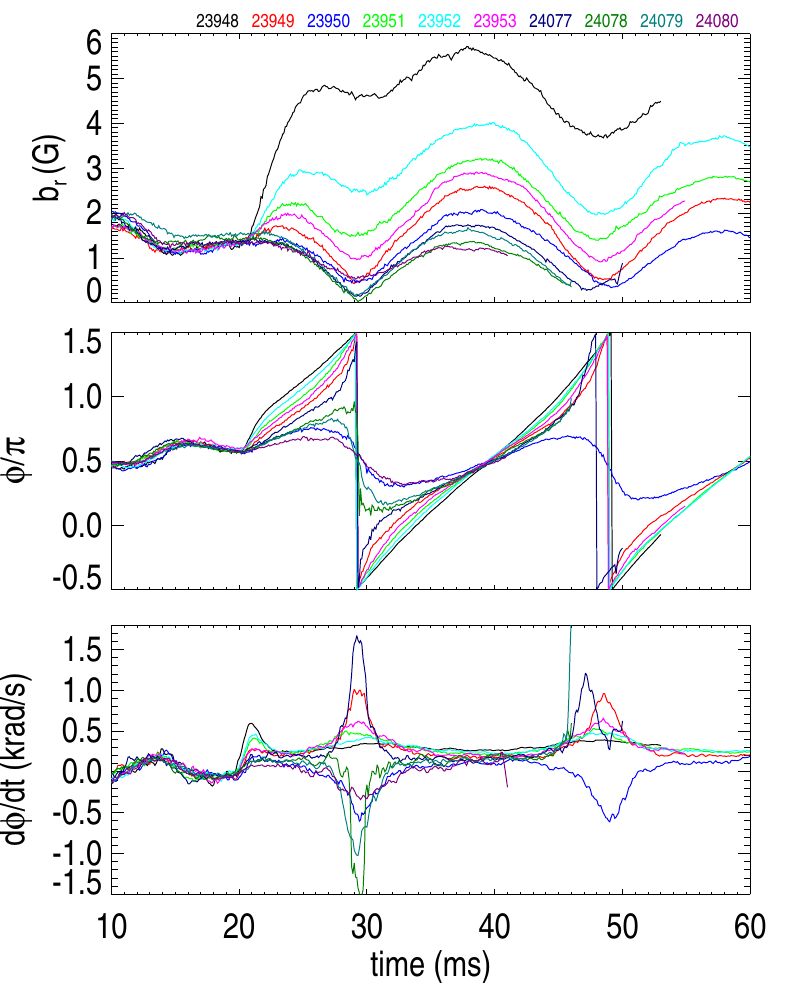}}
    \caption{Like Fig.\ref{Fig_n8LProxyRaw}, 
      except that no ``proxy'' EF is applied. The only 
      static $b_0$ is the intrinsic $m$=1, $n$=8 EF. 
      Transition to complete rotation occurs between $b_1=$0.08mT 
      and $b_1=$0.095mT.}
     \label{Fig_n8IntrinsRaw}
     %%%%%% SHOTS 23948-54 
\end{figure}

\begin{figure}[ht]
    \centerline{\includegraphics{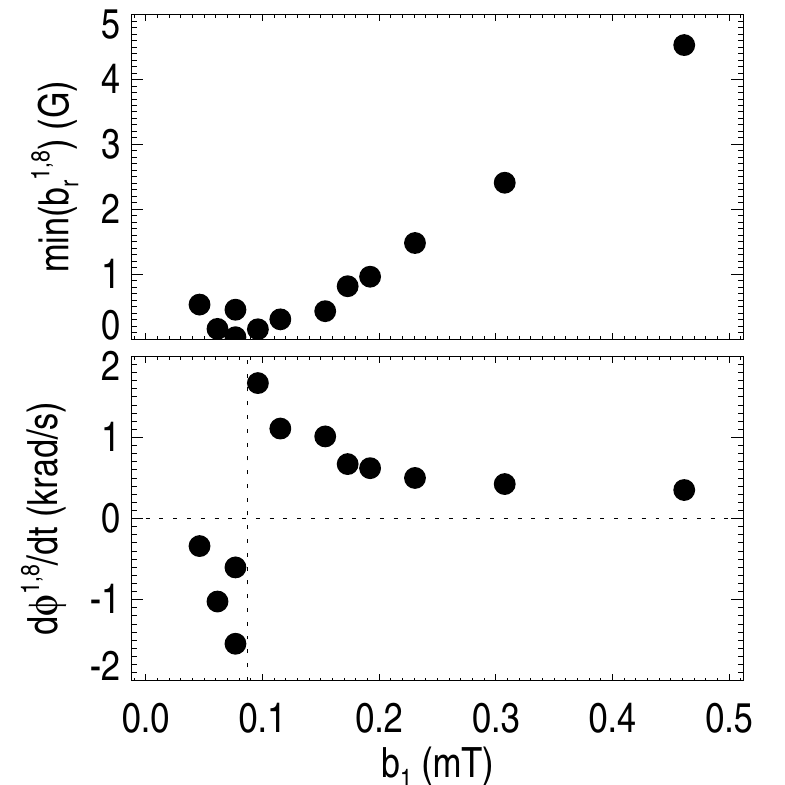}}
  \caption{Like Fig.\ref{Fig_n8LProxyAn}, but for the intrinsic $n$=8 
    results of Fig.\ref{Fig_n8IntrinsRaw}.}
    \label{Fig_n8IntrinsAn}
\end{figure}

\begin{figure}[ht]
    \centerline{\includegraphics{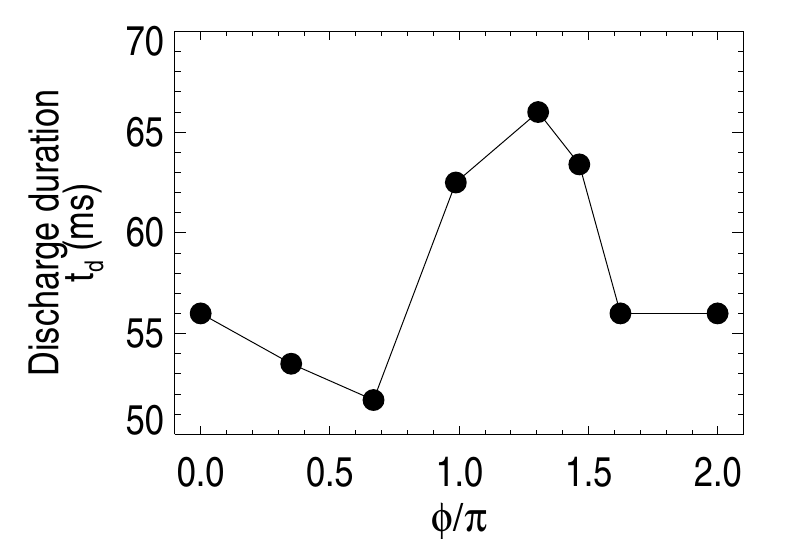}}
    \caption{Optimization of toroidal phase $\phi$ of intrinsic $n$=8 EF 
      correction (EFC). Longer discharge durations are indicative of better 
      EFC.}
    \label{Fig_n8_PhiScan}
\end{figure}

Ultimately the technique was applied in the absence of proxy EFs, to 
assess the intrinsic $m$=1, $n$=8 EF 
(Figs.\ref{Fig_n8IntrinsRaw}-\ref{Fig_n8IntrinsAn}). 
It was established that it has a toroidal phase $\phi/\pi=0.5\pm0.1$ and an 
amplitude of 0.15$\pm$0.02 mT, in agreement with earlier estimates 
\cite{Brunsell,Drake_RWMs}. 
A $\phi$ scan of the discharge duration confirms that the best EFC is 
obtained for a phase $\phi/\pi=1.3\pm0.2$ (Fig.\ref{Fig_n8_PhiScan}), 
which is opposite to the EF phase $\phi/\pi=0.5$, as expected. 

With this information taken into account, it is deduced 
that the vector sum of the intrinsic EF and applied static MP in the 
$n$=8 proxy experiment described above had  
amplitude $b_0=0.44$mT, agreeing within two standard 
deviations with the measurement $b_0=0.77\pm0.08$mT, and orientation 
$\phi/\pi =0.11\pm0.1$, 
in good agreement with the measurement, $\phi/\pi=0.1\pm0.1$.

%===========================================================================
\section*{Summary and Discussion}
In summary a new non-disruptive error field (EF) assessment technique
not  restricted to low density and thus low beta was demonstrated at
the  EXTRAP-T2R reversed field pinch. Stable Resistive Wall Modes (RWMs) 
of toroidal mode number $n$=8 and 10 were
generated and their rotation sustained by rotating magnetic
perturbations.  
Due to finite EFs, and in spite of the applied perturbations rotating
uniformly and having constant amplitude, the RWMs were observed to
rotate non-uniformly and be modulated in amplitude 
(Figs.\ref{Fig_n10LProxyRaw}, \ref{Fig_n10LProxyAn}, 
\ref{Fig_n10IntrinsRaw}, \ref{Fig_n10IntrinsAn}, 
\ref{Fig_n8LProxyRaw}, \ref{Fig_n8LProxyAn}, 
\ref{Fig_n8IntrinsRaw}, \ref{Fig_n8IntrinsAn},). 
This behavior was interpreted with a simple theoretical model 
(Eq.\ref{Eq_bsim}) and 
used to characterize and correct intrinsic $n$=8 and 10 EFs, leading to  
longer discharges and more uniform mode rotation 
(Figs.\ref{Fig_n10_PhiScan}, \ref{Fig_n8_PhiScan}). 

The probing perturbation was rotated (i.e., its phase was scanned) during the 
discharge, while its amplitude was scanned from shot to shot. 
It is conceivable that in another device with sufficiently long discharges,  
such as ITER, 
both parameters (amplitude and phase) can be scanned within a single 
discharge or even a fraction of it, e.g.~by deploying a rotating, 
growing (``spiraling'') probing field. While fast relative to the 
discharge duration (to guarantee 
several rotation periods and a thorough scan), for ease of interpretation 
the MP rotation and growth should be slow compared with the RWM growth, the 
resistive wall time  and current diffusion time. 

The technique works best with non-growing modes driven by or sensitive to EFs 
(or, more precisely, to the EF+MP resultant). The principle is that when the 
drive is reduced or rotated, the mode shrinks or rotates accordingly. 
At EXTRAP-T2R we used stable 
Resistive Wall Modes (RWMs) of $n$=8 and 10 for the reason that these modes, 
indeed, do not grow. There was no fundamental reason behind the high $n$. 

The technique is easily 
extended to classical or neoclassical tearing modes in other 
devices, e.g.~$n$=1 locked modes at DIII-D, due to EF penetration or not, 
provided they naturally 
saturate, as it is often the case for periods of several hundreds of ms, 
or are kept small by stabilizing Electron Cyclotron (EC) Heating. 
EC Current Drive is even more stabilizing but its effect depends on its 
toroidal phase relative to the island O-point and, in fact, can become 
destabilizing if deposited in the X-point. This can be used to the 
advantage of the technique presented, as it enhances the variations of the 
measured field, but attention has to be paid to separate the effects of the 
EF and of the current drive. 

The extension to unstable modes such as $n=-1$ RWMs at EXTRAP-T2R is 
left as future work.

\section*{Acknowledgments}
This work was partly supported by U.S. DOE Contract DE-SC0008520.

%===========================================================================
\section*{References}
%\bibliography{Draft}

%===========================================================================

\clearpage
\newpage

% n=10, large proxy ===================================================

% n=10, intrinsic ======================================================

% n=7, 8, 9, 10 ======================================================

% n=8, proxy ========================================================

% n=8, intrinsic  ========================================================

\end{document}